\documentclass[epj]{svjour}
\usepackage{graphics}
\begin{document}
\title{The New Narrow ``$D_s$'' States -- A Minireview}
\author{Frank C. Porter%
}
\institute{(From the BaBar collaboration)\hfil\break Lauritsen Laboratory of Physics 356-48, California Institute of Technology, Pasadena, CA 91125 \\ \\
To appear in Proceedings of the International Europhysics Conference on High Energy Physics,
July 17-23, 2003, Aachen}
\date{}
\abstract{The experimental status concerning the two new narrow states with charm-strange content is reviewed. The states have masses of
2317 and 2460 MeV, widths less than 10 MeV, isospin consistent with zero, and spin-parities consistent with being $0^+$ and $1^+$, respectively. Although the masses are lower than the conventional expectation, these states appear to be the $j=1/2$ P-wave levels of the $D_s$ system, where $j$ is the light quark angular momentum; there may be mixing with the $j=3/2$ level for the $1^+$ state. 
\PACS{
      {14.40.Lb}{Charmed mesons} \and
      {13.66.Bc}{Hadron production in e-e+ interactions}
     } 
} 
\maketitle
\section{Introduction}
\label{sec:intro}

Two new narrow states have been observed with charm and strangeness.
The current state of experimental knowledge is summarized, drawing experimental results from
BaBar\cite{bib:BaBar}, Belle\cite{bib:Belle}, CDF\cite{bib:CDF}, and CLEO\cite{bib:CLEO}.
These new states will be referred to here as $D^*_{sJ}(2317)^\pm$ and $D_{sJ}(2460)^\pm$, in accordance with Particle Data Group naming conventions and their presumed quantum numbers in the $q\bar q$ model.

\section{The $D_{sJ}^*(2317)^\pm$ Level}
\label{sec:Ds2317}

\begin{figure}
\resizebox{0.5\textwidth}{!}{%
 \vbox{
  \hbox{
    \includegraphics{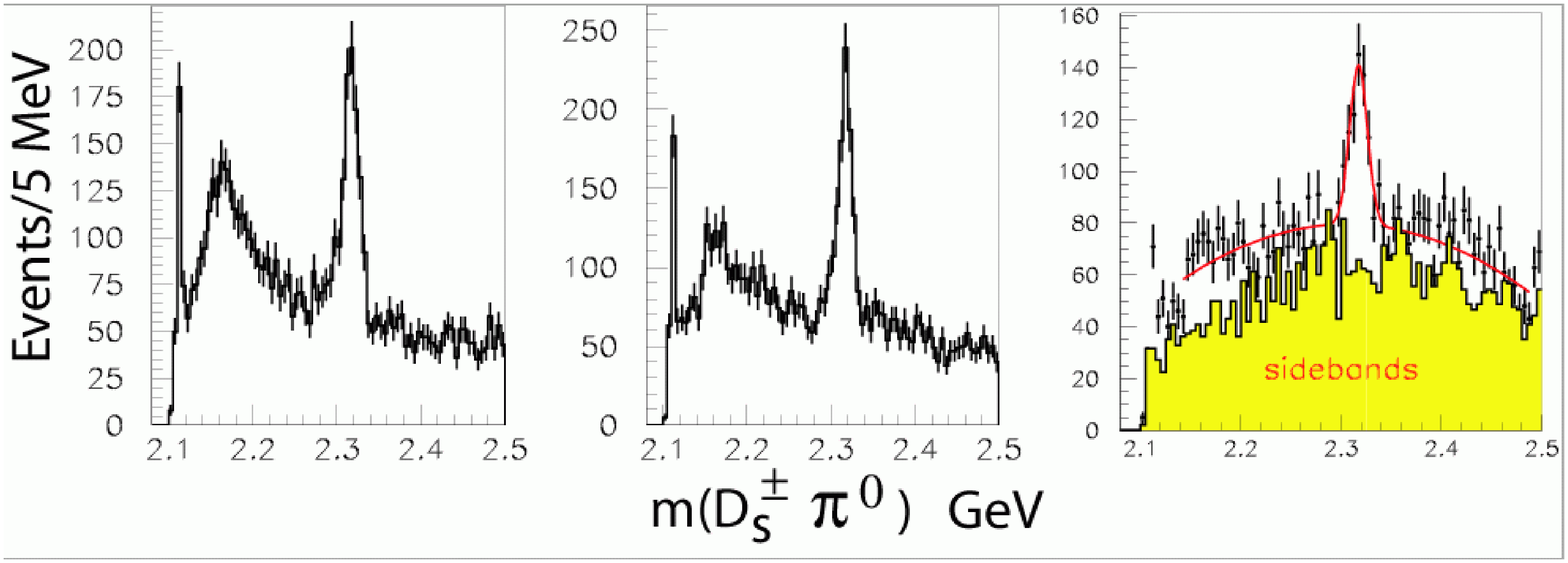}
  }\hbox{
   \resizebox{37.5cm}{!}{
    \includegraphics{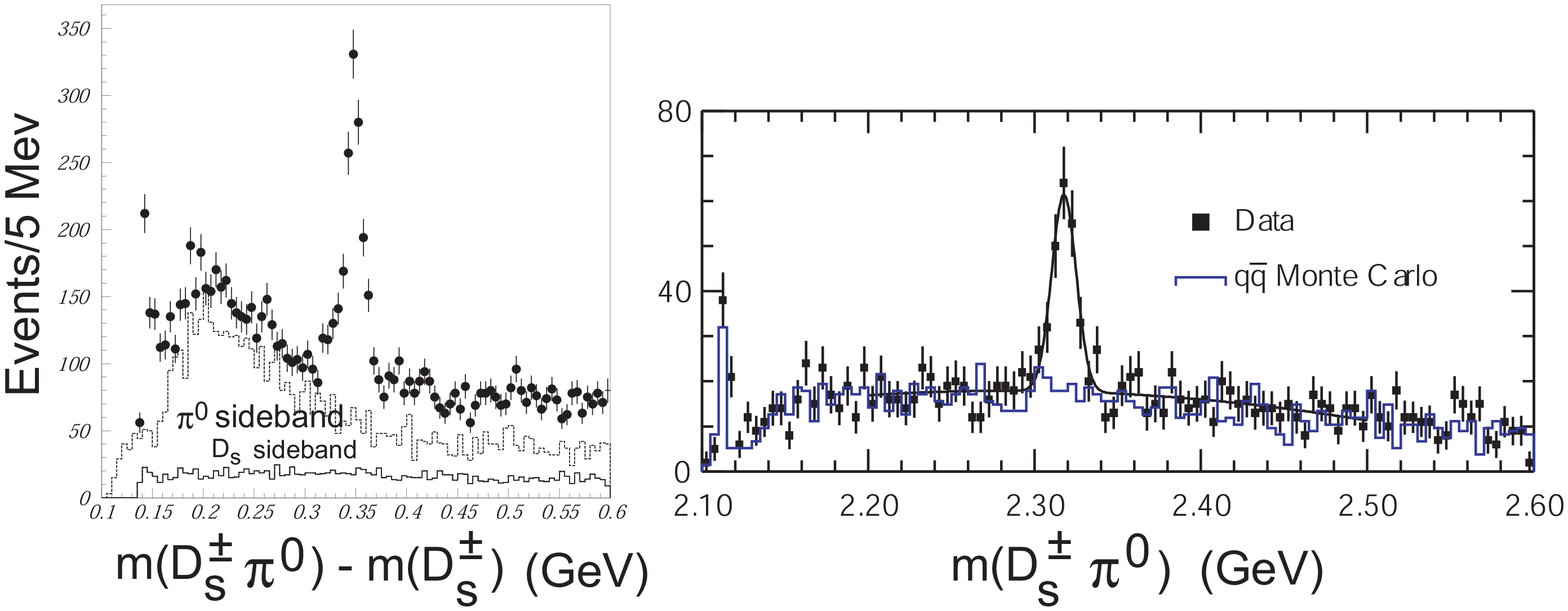}}
  }
 }
}
\caption{The $D_s^\pm\pi^0$ mass spectrum, for $p^*>3.5$ GeV. Top row: BaBar, with $D_s^\pm\to \phi\pi^\pm$ (left), $D_s^\pm\to K^{*0}K^\pm$ (middle), $D_s^\pm\to K^+K^-\pi^\pm\pi^0$ (right). Bottom left: Belle preliminary, with $D_s^\pm\pi^0 - D_s^\pm$ mass difference, where $D_s^\pm\to\phi \pi^\pm$; Bottom right: CLEO, with $D_s^\pm\to \phi\pi^\pm$.}\label{fig:2317peak}
\end{figure}

The $D^*_{sJ}(2317)^\pm$ state is observed as a mass peak in the $D_s^\pm\pi^0$ spectrum, Fig.~\ref{fig:2317peak}. A $p^*>3.5$ GeV CM momentum cut ensures that the signal observed here is from continuum charm production (not from $B$ decays). The peak width is consistent with experimental resolution, hence the state is narrow, with a width less than several MeV.

\begin{figure}[t]
\resizebox{0.5\textwidth}{!}{
 \hskip-1cm\includegraphics{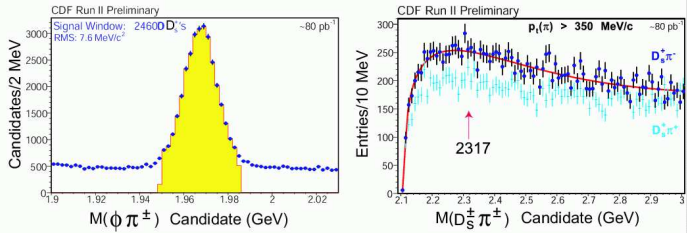}
}\vskip-0cm
\caption{CDF preliminary. Left: The $D_s^\pm$ signal in $\phi\pi^\pm$. Right: Search for a peak at 2317 MeV in the $D_s^\pm\pi^\pm$ spectra, dark points are opposite-sign, light points are same sign.}\label{fig:CDFsearch}
\end{figure}

The narrow width suggests an isospin violating decay, ie $I\ne 1$. CDF has looked for a peak in the $D_s^\pm\pi^\mp$ and $D_s^\pm \pi^\pm$ spectra, Fig.~\ref{fig:CDFsearch}. While they have a large $D_s^\pm$ signal, there is no evidence for a peak in these modes.  The CDF sensitivity is estimated using $D^*_2\to D\pi$ decays and the relative $D_s\!:\!D$ rates. Further work is underway to quantify the sensitivity using $D_{sJ}(2573)\to DK$, but it presently seems that CDF has the sensitivity to observe isospin partners if they exist. The lack of such a signal leads to the preferred $I=0$ assignment. 

\begin{figure}
\resizebox{0.5\textwidth}{!}{%
\includegraphics{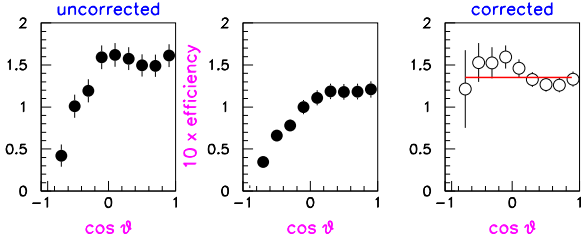}
}
\caption{BaBar helicity angle ($\theta$) analysis of the decay $D^*_{sJ}(2317)^\pm \to D_s^\pm\pi^0$. Left: Uncorrected angular distribution. Center: Dependence of efficiency on angle. Right: Efficiency-corrected angular distribution.}\label{fig:2317helicity}
\end{figure}

\begin{figure}
\resizebox{0.5\textwidth}{!}{%
 \hbox{
  \vbox{\hskip-.8cm
   \resizebox{.80\textwidth}{!}{
   \includegraphics{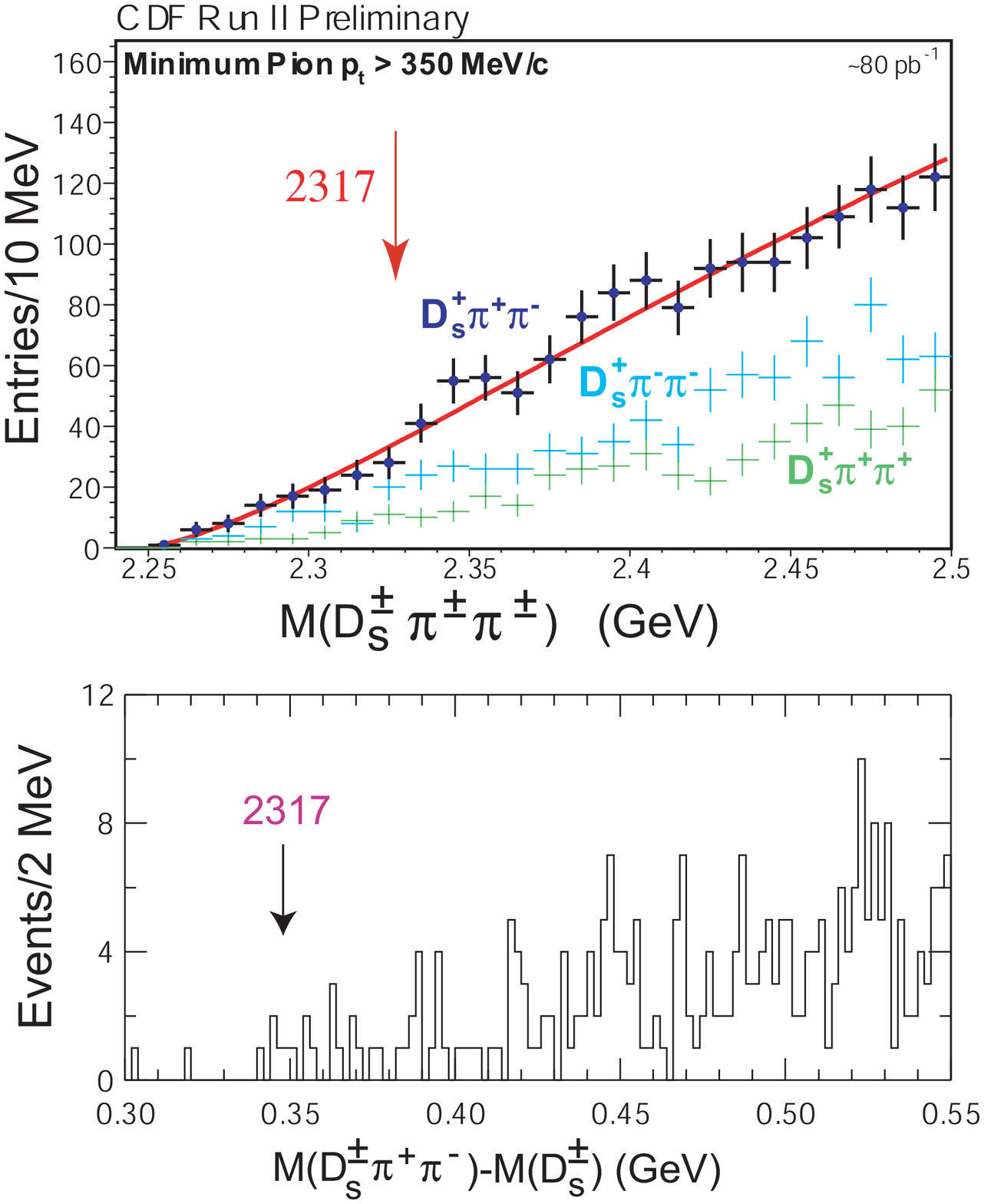}}
  }\raise1cm\vbox{\hskip.8cm
   \resizebox{7.5cm}{!}{
   \includegraphics{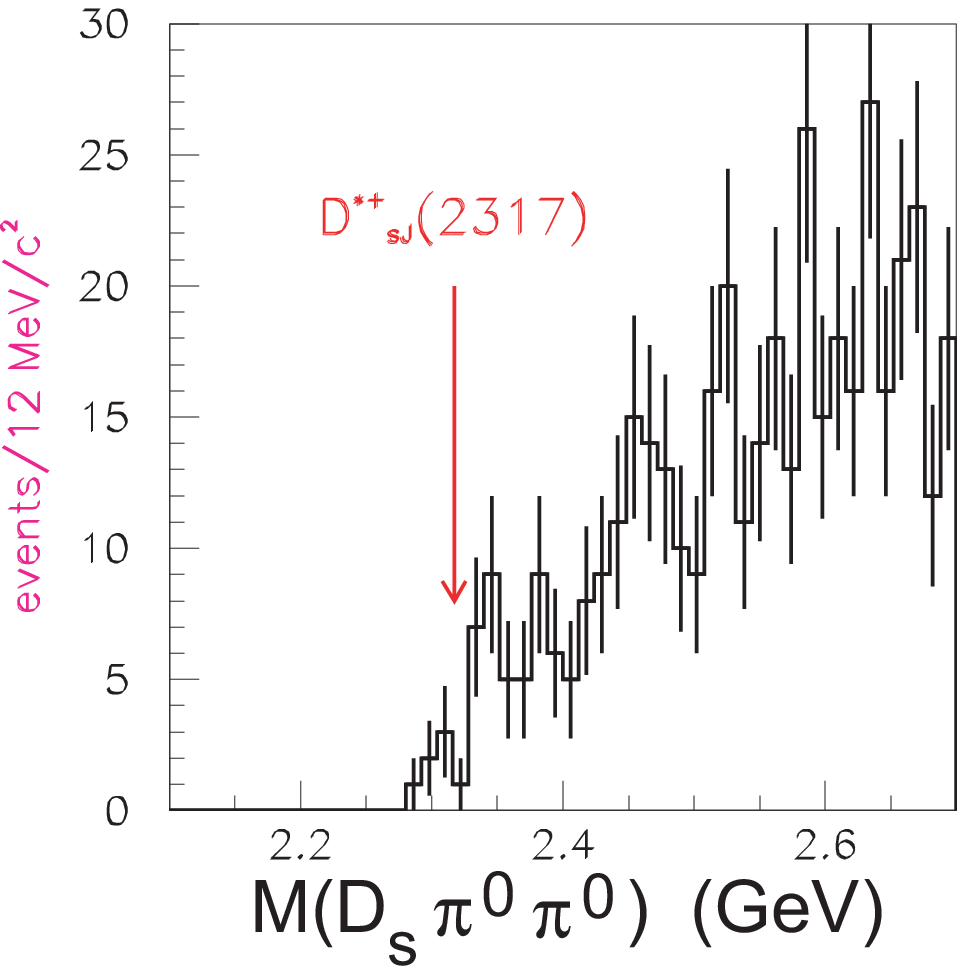}
   }
  }
 }
}
\caption{Search for signals in the $D_s^\pm\pi\pi$ spectra. Arrows indicate where a signal for the $D_{sJ}^*(2317)^\pm$ would peak. Top left: CDF preliminary, dark points are opposite sign pions, middle points are like sign pions, opposite sign with $D_s^\pm$, and lower points are like sign pions, same sign with $D_s^\pm$. Bottom left: CLEO, $D_s^\pm\pi^+\pi^- - D_s^\pm$ mass difference. Right: BaBar, $D_s^\pm\pi^0\pi^0$.}\label{fig:2317pipi}
\end{figure}

\begin{figure}
\resizebox{0.52\textwidth}{!}{%
 \hskip-0.5cm\includegraphics{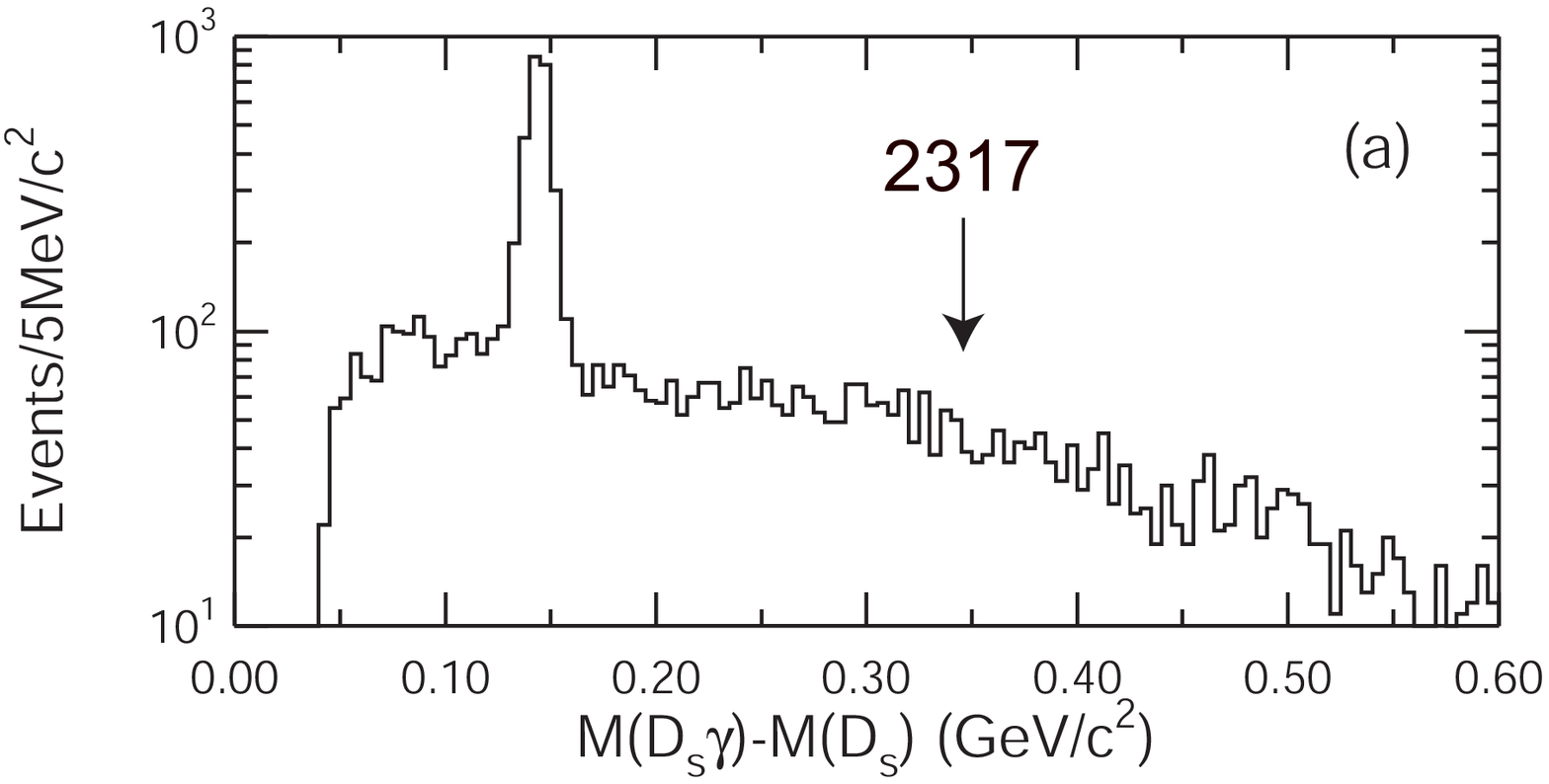}
\includegraphics{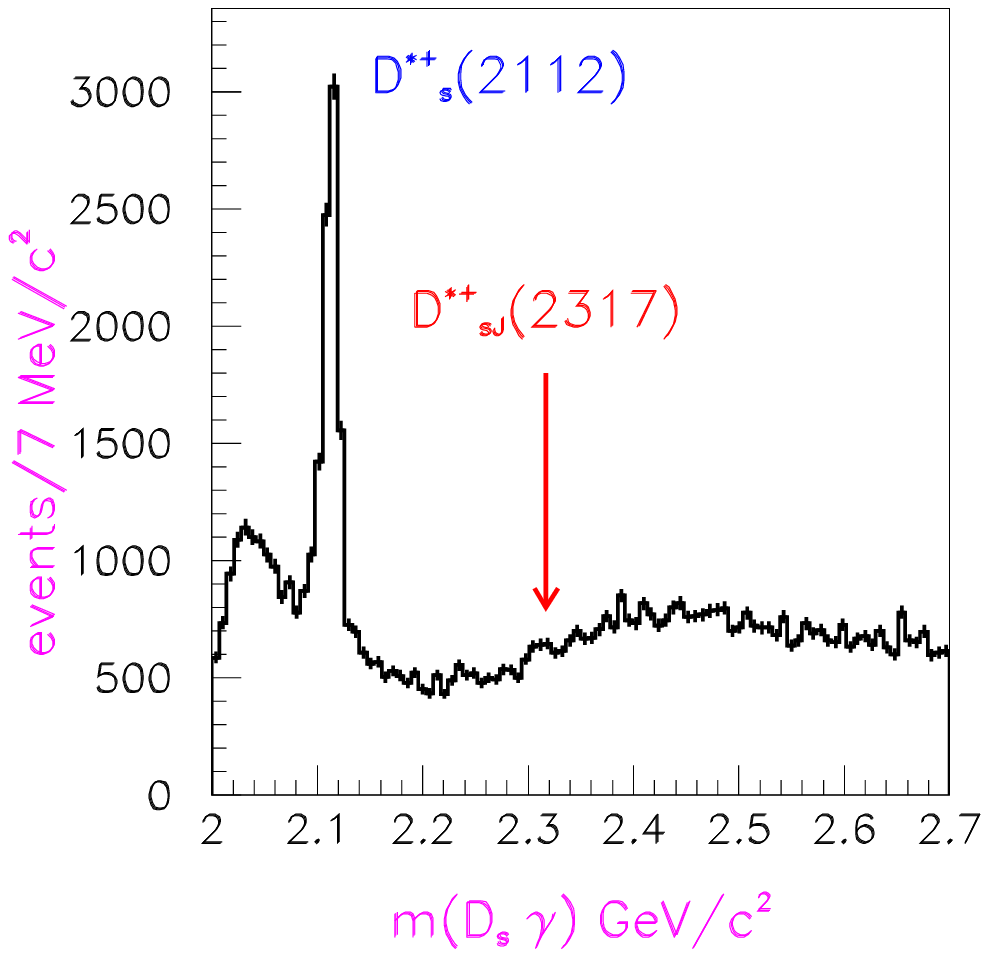}
}
\caption{Search for signals in the $D_s^\pm\gamma$ spectrum. The arrows indicate where a signal for the $D_{sJ}^*(2317)^\pm$ would appear. The peak at lower mass is from $D_s^*(2112)^\pm \to D_s^\pm\gamma$. Left: CLEO, $D_s^\pm\gamma - D_s^\pm$ mass difference. Right: BaBar, $D_s^\pm\gamma$ mass.}\label{fig:2317gamma}
\end{figure}

If parity is conserved, then the decay $D^*_{sJ}(2317)^\pm \to D_s^\pm\pi^0$ implies that $J^P$ is natural for the $D^*_{sJ}(2317)^\pm$ (hence, the $^*$ in its name). There is some evidence favoring a $J^P=0^+$ assignment: First the helicity angle distribution for the decay is consistent with uniform, Fig.~\ref{fig:2317helicity}. This is consistent with $J^P=0^+$, but doesn't rule out other possibilities, because it can also result from an isotropic production polarization. Further support for the $0^+$ hypothesis is gleaned from the absence of a peak in the $D_s^\pm\pi^+\pi^-$ and $D_s^\pm\pi^0\pi^0$ spectra, Fig.~\ref{fig:2317pipi}, forbidden for $0^+$. Furthermore, a spin zero decay to $D_s^\pm\gamma$ is forbidden, and is not observed, Fig.~\ref{fig:2317gamma}. Ultimately, we look forward to an unambiguous spin-parity determination in an angular analysis of $B\to D D^*_{sJ}(2317)^\pm$ decays.

\begin{figure}
\resizebox{0.5\textwidth}{!}{%
\includegraphics{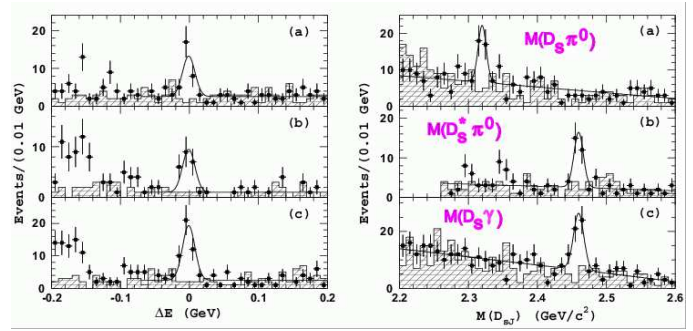}
}
\caption{Belle preliminary analysis of the channels $B\to DD_s\pi^0$ and $DD_s\gamma$, where $D_s^+\to \phi\pi^+$, $\bar K^{*0}K^+$, and $K_S^0K^+$ (and charge conjugate modes) and $D \to K\pi, K\pi\pi$, and $K\pi\pi\pi$, in a sample of $124\times 10^6\,B\bar B$ events. The plots on the left are the $\Delta E \equiv \sum_i\sqrt{m_i^2+{\bf p}_i^{*2}}-E_b^*$ distributions, and on the right the $M_{ES}\equiv\sqrt{E_b^{*2}-|\sum_i{\bf p}_i^*|^2}$ distributions, where $m_i$ and ${\bf p}_i^*$ are the masses and three-momenta of the candidate $B$ decay products in the $\Upsilon(4S)$ frame, and $E_b^*$ is the beam energy in the $\Upsilon(4S)$ frame.}\label{fig:exclusiveB}
\end{figure}

There is hope for such an analysis; the Belle experiment has already observed the $D^*_{sJ}(2317)^\pm$ in exclusive $B\to DD_s\pi^0$ decays, as shown in Fig.~\ref{fig:exclusiveB}a and summarized in Table~\ref{tab:2317B}. As more data is accumulated, an angular analysis will become possible. Table~\ref{tab:2317sum} gives a summary of the other information on the $D^*_{sJ}(2317)^\pm$.

\begin{table}
\caption{Product branching fractions in exclusive $B$ decays to $D_{sJ}^*(2317)^\pm$ or $D_{sJ}(2460)^\pm$ (Belle preliminary). The yields are obtained from fits to the $\Delta E$ distributions (Fig.~\ref{fig:exclusiveB}, left).}
\label{tab:2317B}
\begin{tabular}{lll}
\hline\noalign{\smallskip}
$B$ Decay channel & Yield  & ${\cal B} (10^{-4})$ \\ \hline\noalign{\smallskip}
$\bar D^0D_{sJ}^*(2317) \to \bar D^0 D_s\pi^0$ \hfill & $13.7^{+5.1}_{-4.5}$ & $8.1^{+3.0}_{-2.7}\pm 2.4$ \\
$\bar D^-D_{sJ}^*(2317)\to D^-D_s\pi^0$ \hfill & $10.3^{+3.9}_{-3.1}$ & $8.6^{+3.3}_{-2.6}\pm 2.6$ \\
$\bar D^0D_{sJ}^*(2317)\to \bar D^0D_s\gamma$ \hfill & $3.4^{+2.8}_{-2.2}$ & $2.4^{+2.0}_{-1.5} (<5.7)$ \\
$\bar D^-D_{sJ}^*(2317)\to \bar D^-D_s\gamma$ \hfill & $2.3^{+2.5}_{-1.9}$ & $2.6^{+2.8}_{-2.2} (<7.1)$ \\ 
\noalign{\smallskip}\hline\noalign{\smallskip}
$\bar D^0D_{sJ}(2460)\to \bar D^0D_s^*\pi^0$ \hfill & $7.2^{+3.7}_{-3.0}$ & $11.9^{+6.1}_{-4.9}\pm 3.6$ \\
$D^-D_{sJ}(2460)\to D^-D_s^*\pi^0$ \hfill & $11.8^{+3.8}_{-3.2}$ & $22.7^{+7.3}_{-6.2}\pm 6.8$ \\
$\bar D^0D_{sJ}(2460)\to \bar D^0D_s\gamma$ \hfill & $19.1^{+5.6}_{-5.0}$ & $5.6^{+1.6}_{-1.5}\pm 1.7$ \\
$D^-D_{sJ}(2460)\to D^-D_s\gamma$ \hfill & $18.5^{+5.0}_{-4.3}$ & $8.4^{+2.4}_{-2.2}\pm 2.5$ \\
$\bar D^0D_{sJ}(2460)\to \bar D^0D_s^*\gamma$ \hfill & $4.4^{+3.8}_{-3.3}$ & $3.1^{+2.7}_{-2.3}(<7.5)$ \\ 
$D^-D_{sJ}(2460)\to D^-D_s^*\gamma$ \hfill & $1.1^{+1.8}_{-1.2}$ & $1.3^{+2.0}_{-1.4}(<4.6)$ \\ 
$\bar D^0 D_{sJ}(2460)\to \bar D^0D_s\pi^0$ \hfill & $<2.4$ & $<2.2$ \\ 
$D^- D_{sJ}(2460)\to D^-D_s\pi^0$ \hfill & $<2.4$ & $<2.8$ \\ 
$\bar D^0D_{sJ}(2460)\to \bar D^0D_s\pi^+\pi^-$ \hfill & $<4.0$ & $<2.4$ \\
$D^-D_{sJ}(2460)\to D^-D_s\pi^+\pi^-$ \hfill & $<2.5$ & $<2.0$\\ \hline
\end{tabular}
\end{table}

\begin{table}
\caption{Summary of $D_{sJ}^*(2317)^\pm$. Limits are 90\% C.L. Masses and widths are in MeV. $^\dagger$The Belle peak has $\sigma=7.6\pm0.5$ MeV, consistent with resolution.
}
\label{tab:2317sum}
\resizebox{.5\textwidth}{!}{
\begin{tabular}{l@{\hskip-1cm}ccc}
\hline\noalign{\smallskip}

Quantity                              & {  BaBar}         & {  Belle} & {  CLEO} \\ \hline
Dataset (fb$^{-1}$)                   & 91 & 87 & 13.5 \cr
$D_s$ Modes \hfill & $\phi\pi$, $K^*K$ & $\phi\pi$ & $\phi\pi$ \cr
\smash{\lower5pt\hbox{Mass}}     \hfil             & $2316.8\pm 0.4 \pm 3$    & & $2318.5\pm 1.2\pm 1.1$ \\
 & \multispan{3}{\hfill $2317.2\pm0.5\pm0.9$\hfill} \\
\smash{\lower5pt\hbox{$\Delta m(2317-D_s)$}}  \hfil & $348.4 \pm 0.4 \pm 3$ & & $350.0\pm 1.2\pm 1.0$ \\
 & \multispan{3}{\hfill $348.7\pm 0.5 \pm 0.7$\hfill} \\
Width      \hfil      & $<10$ & {  $\dagger$} & $<7$ \cr
${{\cal B} (D_s\pi^+\pi^-) / {\cal B}(D_s\pi^0)}$ \hfill \hfil & & & $<0.019 $ \\
${{\cal B}(D_s\gamma)  / {\cal B}(D_s\pi^0)}$ \hfill  \hfil && $<0.05$ & $< 0.052$ \\
${{\cal B}(D_s^*\pi^0) / {\cal B}(D_s\pi^0)}$  \hfill  & & & $<0.11 $ \\
${{\cal B}(D_s^*\gamma)  / {\cal B}(D_s\pi^0)}$ \hfill &&& $< 0.059$ \cr
\multispan{2}{\strut ${\sigma\cdot{\cal B}(D_{sJ}^*\to D_s\pi^0) \over \sigma(D_s)}$ ($p^*>3.5$ GeV) \hfill}& & \hskip-1cm $(7.9\pm 1.2\pm 0.4)\times 10^{-2}$ \\ \multispan{4}{\phantom{.}} \\
\hline
\end{tabular}
}
\end{table}

\section{The $D_{sJ}(2460)$ Level}
\label{sec:Ds2460}

\begin{figure}
\resizebox{0.5\textwidth}{!}{%
\resizebox{0.34\textwidth}{!}{%
 \includegraphics{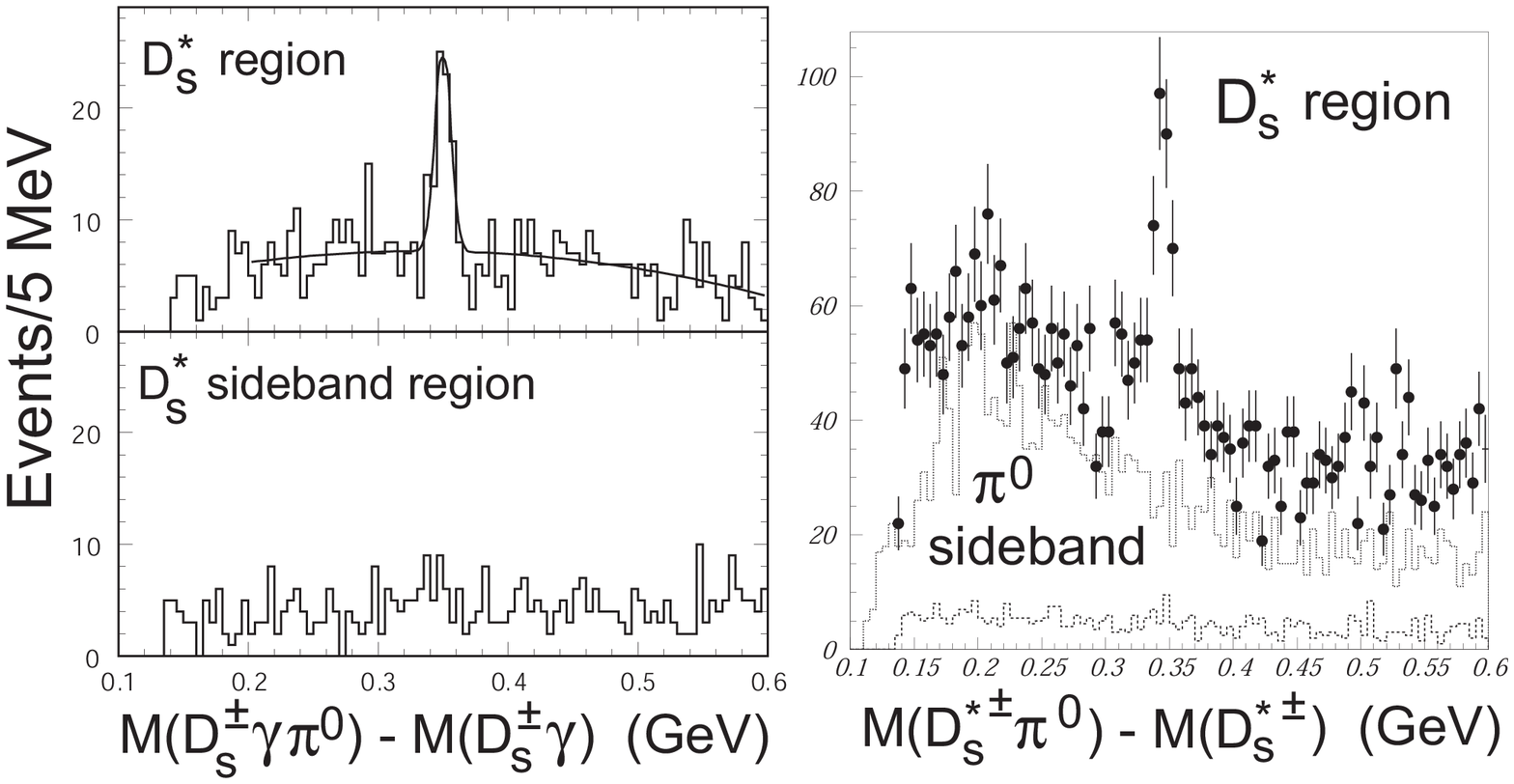}
}
\resizebox{0.16\textwidth}{!}{%
 \raise.5cm\vbox{\includegraphics{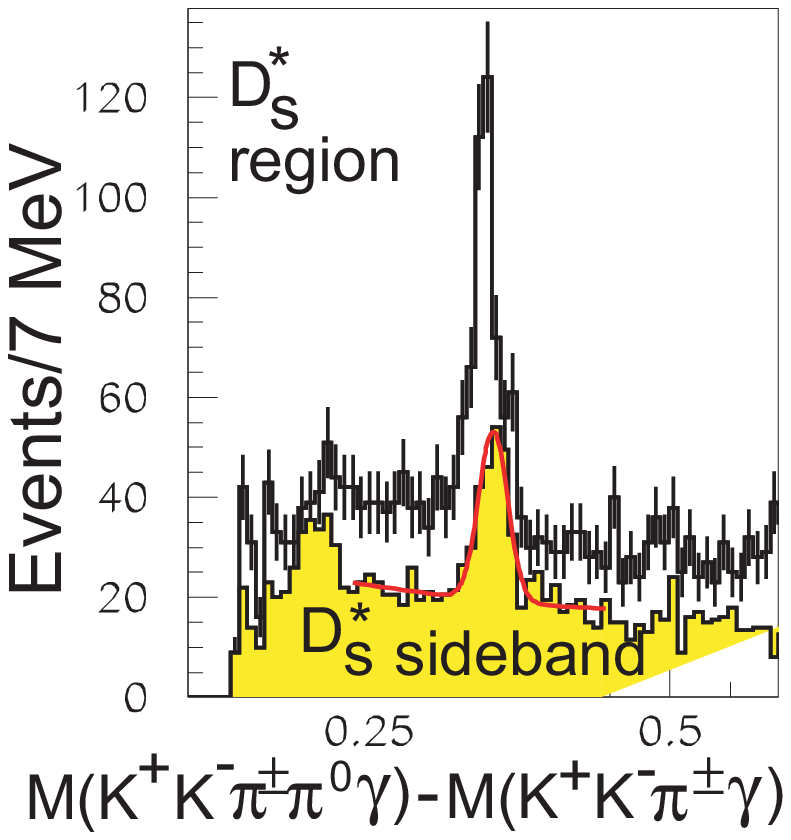}}
}
}
\caption{Observation of the $D_{sJ}(2460)^\pm$ in the $D_s^\pm\gamma\pi^0$ mass spectrum. Left: CLEO. Middle: Belle preliminary. Right: BaBar preliminary.}\label{fig:2460peak}
\end{figure}

The $D_{sJ}(2460)^\pm$ state is observed as a mass peak in the $D_s^\pm\pi^0\gamma$ spectrum, selected on the $D_s^*(2112)^\pm$ region of the $D_s^\pm\gamma$ mass, Fig.~\ref{fig:2460peak}. A $p^*>3.5$ GeV  CM momentum cut ensures that the signal observed here is not in $B$ decays. The peak width is consistent with experimental resolution, hence the state is narrow, with a width less than several MeV. Again, this is indicative of an isospin-violating decay. This state is not seen in the $D_s^\pm\pi^\pm$ and $D_s^\pm\pi^+\pi^-$ spectra, Figs.~\ref{fig:CDFsearch} and \ref{fig:2317pipi}, nor in the $D_s^\pm\pi^+\pi^-$ spectrum in exclusive $B$ decays against a $D$ from Belle (Table~\ref{tab:2317B}).

The $D_{sJ}(2460)$ is above threshold for isospin-allowed decays to $DK$. The fact that this decay is not observed is suggestive that this level has unnatural $J^P$ quantum numbers (hence, no ``$*$'' in the name).

There is a serious difficulty with a cross-feed ambiguity, due to the fact that 
\begin{eqnarray*}
 \Delta m\left[D_{sJ}(2460)-D^*_s(2112)\right] &\sim& \Delta m\left[D_{sJ}^*(2317)-D_s\right] \\
  &\sim& 350\,\hbox{MeV}.
\end{eqnarray*} 
This coincidence has two potential consequences: First, the $D^*_{sJ}(2317)$ signal could feed up to create a peak at $\sim2460$ by adding a random photon selected to be consistent with the decay $D^*_s(2112)\to D_s\gamma$. Second, a signal for $D_{sJ}(2460)$ could create a peak at $\sim 2317$ by neglecting the photon in the $D^*_s(2112)\to D_s\gamma$ transition. There is also an ambiguity in the $D_{sJ}(2460)$ decays: Is it $D_{sJ}(2460)\to D^*_s(2112)\pi^0$, or is it $D_{sJ}(2460)\to D^*_{sJ}(2317)\gamma$? All three experiments which have observed this state correct for the cross-feed between 2317 and 2460 peaks under the $D_{sJ}(2460)\to D^*_s(2112)\pi^0$ hypothesis.

\begin{figure}
\resizebox{0.5\textwidth}{!}{%
\includegraphics{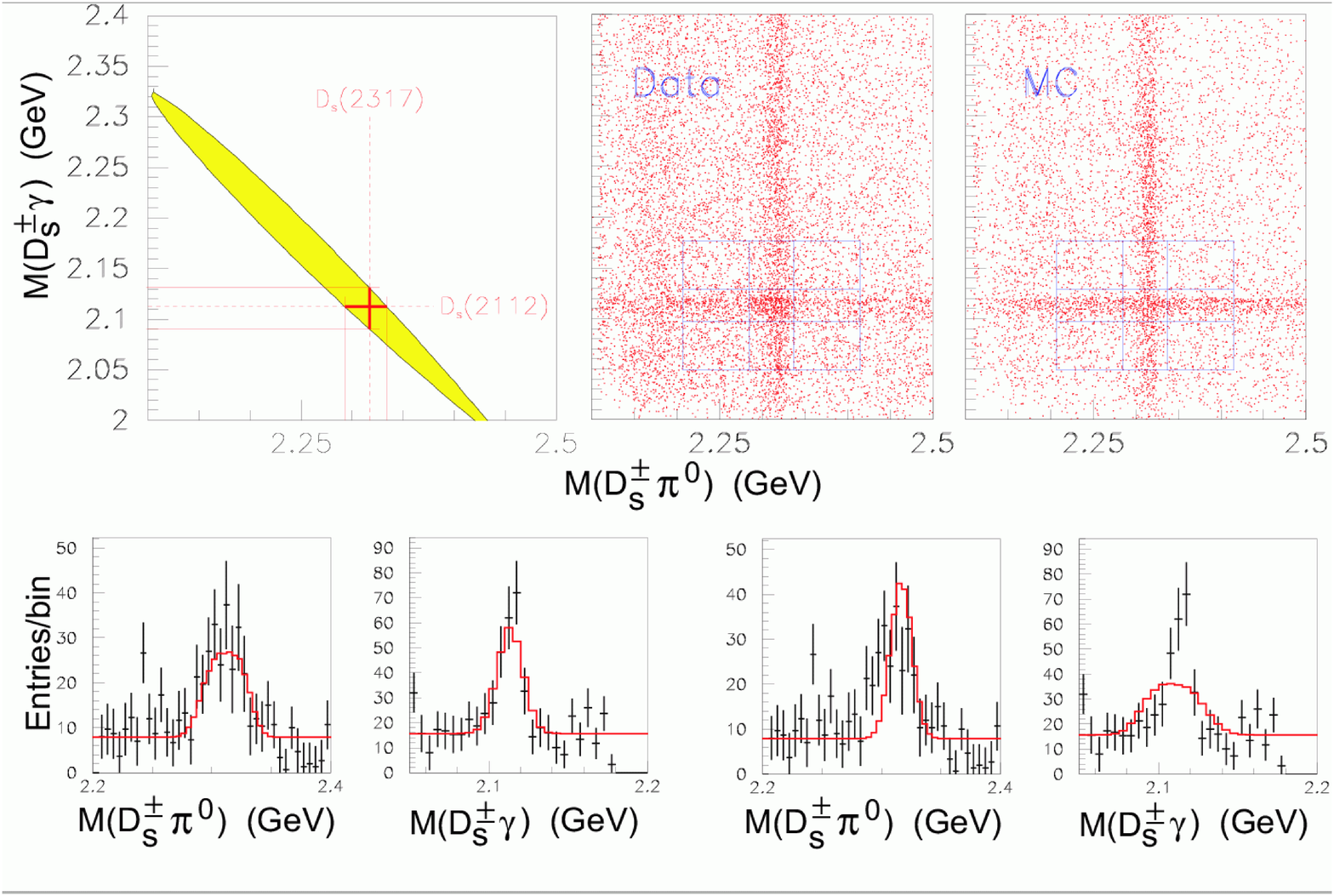}
}
\caption{BaBar preliminary study of the cross-feed between $D_{sJ}(2460)\to D^*_s(2112)\pi^0$ and $D_{sJ}(2460)\to D^*_{sJ}(2317)\gamma$. Top row: plots of $m(D_s^\pm\gamma)$ vs. $m(D_s^\pm\pi^0)$. Left: Kinematic boundary for mass 2460 to $D_s^\pm\pi^0\gamma$. Middle: Data. Right: Monte Carlo simulation for the $D_{sJ}(2460)\to D^*_s(2112)\pi^0$ hypothesis. Bottom row: Projections from the top-middle scatter plot (points with error bars). A linear background subtraction has been performed using the sidebands indicated with the boxes in the top-middle plot. Left: Projections overlaid with histograms according to a simulation of the $D_{sJ}(2460)\to D^*_s(2112)\pi^0$ hypothesis. Right: Projections overlaid with histograms according to a simulation of the $D_{sJ}(2460)\to D^*_{sJ}(2317)\gamma$ hypothesis.}
\label{fig:dalitzStudy}
\end{figure}

BaBar and CLEO have both attempted to test the hypothesis, with consistent results; we use the BaBar study here to illustrate.  Fig.~\ref{fig:dalitzStudy} shows the kinematic boundary for a mass 2460 MeV particle in the $m(D_s\gamma)$ vs $m(D_s\pi^0)$ plane. We see the problem immediately: The two hypotheses correspond to vertical and horizontal lines crossing within the narrow kinematic boundary. The scatter plot of the data is also shown, and for comparison a Monte Carlo sample under the  $D_{sJ}(2460)\to D^*_s(2112)\pi^0$ hypothesis. Because of the experimental resolution, the favored hypothesis is not obvious from the scatter plot, although the band in the overlap region perhaps looks more ``horizontal'' than ``vertical'' However, we may make projections of this scatterplot, with a background subtraction performed in the linear approximation. The result is shown in Fig.~\ref{fig:dalitzStudy}, where the projection plots are shown with curves overlain for the two hypotheses. It may be observed that the $D_{sJ}(2460)\to D^*_s(2112)\pi^0$ hypothesis gives a better fit. The data are consistent with being 100\% due to this channel, though a significant contribution from the other hypothesis cannot yet be excluded.

\begin{figure}
\hskip1.5cm\resizebox{0.3\textwidth}{!}{%
\includegraphics{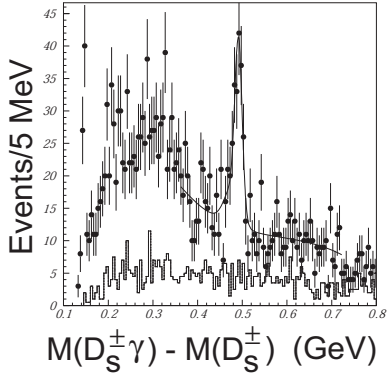}
}
\caption{Belle preliminary observation of $D_{sJ}(2460)^\pm\to D_s^\pm\gamma$ in continuum $e^+e^-$ annihilation. The fit gives $N\left[D_{sJ}(2460)^\pm \to D_s\gamma\right] = 152\pm 18 \pm 11$ signal events. Note that no signal appears at a mass difference around 350 MeV, which would correspond to the 2317 state, consistent with the favored $J=0$ hypothesis for the $D_{sJ}^*(2317)^\pm$. The histogram shows the $D_s^\pm$ sidebands.}
\label{fig:BelleDsgamma}
\end{figure}

\begin{figure}
\resizebox{0.35\textwidth}{!}{%
\hskip2cm\includegraphics{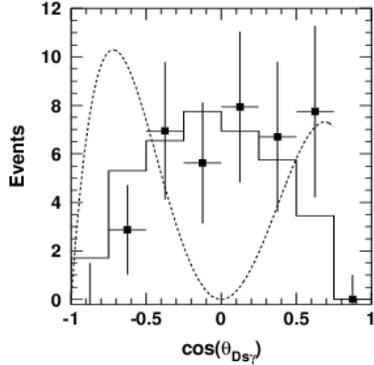}
}
\caption{Cosine of the helicity angle for $D_{sJ}(2460)\to D_s\gamma$ in exclusive $B \to DD_{sJ}(2460)$ decays (Belle preliminary). The data are the points with error bars; the predicton for the spin 1 hypothesis is shown by the histogram, and for the spin two hypothesis by the curve.}
\label{fig:2460BelleAngle}
\end{figure}

The state at 2460 is observed by Belle to decay to $D_s\gamma$, Fig.~\ref{fig:BelleDsgamma}. This rules out $J=0$ for this state. A BaBar angular analysis of $D_{sJ}(2460)$ decays (not shown)
also finds that spin 0 is disfavored. Belle also observes this state in exclusive $B\to D D_{sJ}(2460)$ decays, with $D_{sJ}(2460) \to D_s\gamma$, Fig.~\ref{fig:exclusiveB}c. The product branching fraction is\hfil\break
 ${\cal B}\left[B \to DD_{sJ}(2460) \to DD_s\gamma\right] = \left(6.7^{+1.3}_{-1.2}\pm2.0\right)\times 10^{-4}$. Together with the
observation of $D_{sJ}(2460) \to D_s^*\pi^0$, Fig.~\ref{fig:exclusiveB}b, a branching ratio of $0.38\pm 0.11\pm 0.04$ is obtained for ${\cal B}\left[D_{sJ}(2460) \to D_s\gamma\right]/{\cal B}\left[D_{sJ}(2460) \to D_s^*\pi^0\right]$.
An angular analysis is shown in Fig.~\ref{fig:2460BelleAngle}, finding consistency for $J^P=1^+$ and ruling out spin 2.

A summary of the $D_{sJ}(2460)^\pm$ as seen in exclusive $B$ decays is included in Table~\ref{tab:2317B}, a summary of other properties is shown in Table~\ref{tab:ds2460}.

\begin{table}
\caption{Summary of $D_{sJ}(2460)$. $N$ is the number of events in the peak at 2460 MeV in the $D_s^*(2112)^\pm\pi^0$ spectrum, where $D_s^*(2112)^\pm\to D_s^\pm\gamma$. Limits are 90\% C.L. Masses and widths are in MeV. $^\dagger$Width of mass peak is consistent with resolution.}
\label{tab:ds2460}
\begin{tabular}{l@{\hskip-0.1cm}c@{\hskip-.5cm}c@{\hskip-0.5cm}c}
\hline\noalign{\smallskip}
Quantity                                       & {BaBar}    & {Belle} & {CLEO} \\ \hline
$N$     & $127\pm 22$ & $126\pm 25 \pm 24$ & $41\pm 12$ \\
\smash{\lower5pt\hbox{Mass}}                & $2457.0\pm 1.4 \pm 3$ & & $ 2463.6 \pm 1.7 \pm 1.2$ \\
 & \multispan{3}{\hfill $2456.5\pm 1.3\pm1.1$ \hfill} \\
\smash{\lower6pt\hbox{$\Delta m(2460-D_s^*)$}} & $344.6\pm 1.2 \pm 3$ &  & $351.2\pm 1.7 \pm 1.0$ \\
 & \multispan{3}{\hfill $344.1\pm 1.3 \pm 0.9$ \hfill} \\
Width   \hfil  & {  $\dagger$} & {  $\dagger$} & $<7$ \cr
${{\cal B}(D_s\pi^+\pi^-) \over {\cal B}(D_s^*\pi^0)}$ &&& $<0.08 $ \\
${{\cal B}(D_s\gamma)  \over {\cal B}(D_s^*\pi^0)}$  &&\hfill$0.63\pm 0.15\pm 0.15$ & $< 0.49$ \\
${{\cal B}(D_s^*\gamma) \over {\cal B}(D_s^*\pi^0)}$ &&& $<0.16$ \\
${{\cal B}(D_{sJ}(2317)^*\gamma)  \over {\cal B}(D_s^*\pi^0)\phantom{_Q}}$ &&& $< 0.58$ \\
\hline
\end{tabular}
\end{table}

\section{Production Rates}
\label{sec:ProdRates}

For $p^* > 3.5$ GeV, CLEO has measured production (times decay) rates for the 2317 and 2460 states, normalized to $D_s^\pm$ production:
\begin{eqnarray*}
{\sigma\cdot{\cal B}(D_{sJ}^{*}(2112)^+\to D_s^+\gamma) \over \sigma(D_s^+)} &=& 0.59\pm 0.03\pm 0.01\\
{\sigma\cdot{\cal B}(D_{sJ}^*(2317)^+\to D_s^+\pi^0) \over \sigma(D_s^+)} &=& (7.9\pm 1.2\pm 0.4)\times 10^{-2}\\
{\sigma\cdot{\cal B}(D_{sJ}(2460)^+\to D_s^{*+}\pi^0) \over \sigma(D_s^+)} &=& (3.5\pm 0.9\pm 0.2)\times 10^{-2}.\\
\end{eqnarray*}
Belle (preliminary) has measured the ratio of 2460 to 2317 production (times decay) rates:
\begin{eqnarray*}
{\sigma\cdot{\cal B}(D_{sJ}(2460)^+\to D_s^{*+}\pi^0)\over \sigma\cdot{\cal B}(D_{sJ}^*(2317)^+\to D_s^{+}\pi^0)} &=& 0.26\pm0.05\pm0.06\\
{\sigma\cdot{\cal B}(D_{sJ}(2460)^+\to D_s^{+}\pi^0)\over \sigma\cdot{\cal B}(D_{sJ}^*(2317)^+\to D_s^{+}\pi^0)} &<& 0.06\ (90\% \hbox{ C.L.}).\\
\end{eqnarray*}
The results from the two experiments are consistent.

\section{Conclusions}
\label{sec:Conclusions}

Two new narrow states with $c\bar s$ content, denoted $D_{sJ}^*(2317)^\pm$ and $D_{sJ}(2460)^\pm$, have been observed in $e^+e^-$ collisions. They are consistent with being $0^+$ and $1^+$ states, respectively, and isospin singlets. They are candidates for the two heretofore unseen $P$-wave $j=1/2$ (light quark angular momentum) states of the $D_s$ system, possibly with some mixing  with $j=3/2$ for the $1^+$ state. They are below threshold for isospin-allowed decays to $D^0K^\pm$ or $D^{0*}K$, respectively, in contrast to the consensus expectation. The results from the different experiments are generally consistent; a potential concern is the $D_{sJ}(2460)^\pm$ mass, for which the combined average mass difference (with the $D_s^*(2112)^\pm$) is $346.6\pm 1.2$ MeV, with a $\chi^2$ consistency of $2$\%. 

I am grateful to the BaBar, Belle, CDF, and CLEO collaborations for sharing and discussing their results. This work was supported in part by
DOE grant DE-FG03-92-ER40701.

\end{document}